\documentclass{nature}

\usepackage{graphicx}
\title{Strong-field Coherent Control of Isolated Attosecond Pulse Generation}

\author{Yudong Yang$^{1,2,4}$, Roland E. Mainz$^{1,2,4}$, Giulio Maria Rossi$^{1,2,4}$, Fabian Scheiba$^{1,2}$, \\ Miguel A. Silva-Toledo$^{1,2}$, Phillip D. Keathley$^{3}$, Giovanni Cirmi$^{1,2}$, and Franz X. K\"artner$^{1,2\ast}$}

\begin{document}
\maketitle
\begin{affiliations}
\item Center for Free-Electron Laser Science, Deutsches Elektronen-Synchrotron DESY, Notkestraße 85,
22607 Hamburg, Germany
\item Physics Department and The Hamburg Centre for Ultrafast Imaging, University of Hamburg, Luruper Chaussee 149, 22761 Hamburg, Germany
 \item Research Laboratory of Electronics, Massachusetts Institute of Technology, 77 Massachusetts Ave, Cambridge, 02139 MA, USA
\item These authors contributed equally: Yudong Yang, Roland E. Mainz
and Giulio Maria Rossi.
 \newline
 $\ast \textit{email: franz.kaertner@desy.de}$
\end{affiliations}

\begin{abstract}
Attosecond science promises to reveal the most fundamental electronic dynamics occurring in matter and it can develop further by meeting two linked technological goals related to high-order harmonic sources: higher photon flux (permitting to measure low cross-section processes) and improved spectral tunability (allowing selectivity in addressing specific electronic transitions). New developments come through parametric waveform synthesis, which provides control over the shape of high-energy electric field transients, enabling the creation of highly-tunable isolated attosecond pulses via high-harmonic generation. Here we show that central energy, spectral bandwidth/shape and temporal duration of the attosecond pulses can be controlled by shaping the laser pulse waveform via two key parameters: the relative-phase between two halves of the multi-octave spanning optical spectrum, and the overall carrier-envelope phase. These results not only promise to expand the experimental possibilities in attosecond science, but also demonstrate coherent strong-field control of free-electron trajectories using tailored optical waveforms. 
\end{abstract} % currently 147 words
\newpage
\section*{}
Attosecond pulse generation via high-order harmonic generation (HHG)\cite{mcpherson1987studies,ferray1988multiple} is undoubtedly a milestone in the development of laser technology which elucidated fundamental mechanisms in light-matter interactions\cite{krausz2009attosecond}. Over the last three decades, this technological breakthrough, enabled the exploration of charge dynamics occurring in atoms, molecules and solids on their natural sub-femtosecond timescales. A few examples include real-time observation of valence electron motion in atoms, electron charge migration in molecules and ionization-induced attosecond time delays\cite{drescher2002time,goulielmakis2010real,kraus2015measurement,calegari2014ultrafast,wirth2011synthesized,calegari2016advances,nisoli2017attosecond,schultze2013controlling}.\\
To date, attosecond pulses (isolated or pulse trains) have been generated either by means of HHG\cite{sansone2006isolated,ghimire2019high, luu2018extreme, chen2014generation} or free-electron lasers (FELs)\cite{duris2020tunable, maroju2020attosecond}. Currently the majority of attosecond science experiments are performed with HHG-based sources due to their broad availability in university-scale laboratories. Moreover HHG-based attosecond pulses are intrinsically synchronized to the pump laser and exhibit lower intensity fluctuations than FEL-based sources.\\
For HHG\cite{kuchiev1987atomic,corkum1993plasma,lewenstein1994theory} in gaseous media, an intense laser field first ionizes the atoms and then coherently drives the motion of the liberated electrons. Once the electrons recombine with the parent ions, high-harmonic photons are generated up to the extreme ultraviolet (XUV) or even the soft X-ray spectral region\cite{chang1997generation,popmintchev2012bright}. The entire HHG process takes place within the driving laser optical cycle and repeats every half cycle, leading to a train of high harmonic bursts each with a sub-femtosecond pulse duration. In order to isolate a single HHG event and to obtain an isolated attosecond pulse (IAP), different gating-techniques were developed\cite{chang2007controlling,abel2009isolated,hammond2016attosecond}. The direct generation of IAPs by optical drivers with sub-cycle durations was more recently demonstrated \cite{rossi2020sub}.\\
Despite these important achievements, contemporary attosecond sources impose considerable constraints on experiments, mainly due to the relatively low photon-flux, that prevents the observation of low-cross section processes and complicates the use of low-density samples. The low flux also results in limited spectral tunability achieved by conventional HHG sources, since the use of either a monochromator or an absorbing bandpass filter is in many cases impractical as it would further reduce the photon-flux. This negatively impacts the possibility of selectively addressing specific electronic transitions in atoms and molecules. Solving these problems will improve the applicability of HHG-based sources to a broader range of attosecond science experiments.\\
To significantly increase the possibility of tunability of the attosecond pulses though, a possible path lays inside the HHG process itself. The electric field-dependent nature of ionization, electron excursion, and therefore, also electron recombination enables the control of the HHG-emission characteristics via shaping of the driving optical waveform. Previous studies already suggested that specific optical waveforms can greatly enhance the HHG conversion efficiency, therefore enabling higher photon-flux\cite{chipperfield2009ideal,jin2014waveforms}. The possibility of modifying the attosecond pulse central energy, spectral shape/bandwidth and pulse duration by controlling the driver waveform with sub-cycle resolution remains though experimentally unexplored.\\
Here we report on the generation of tunable IAPs whose spectral-temporal characteristics can be coherently controlled via parametric waveform synthesis, which allows for shaping the HHG driving optical field on a sub-cycle time scale. Differently from previous studies\cite{timmers2016polarization}, that focused on tuning the central photon-energy via macroscopic HHG parameters, such as gas pressure, gas cell position or driving pulse energy, here we concentrate on the control attainable by shaping the driving waveform. The tailored waveform with octave-spanning bandwidth is obtained via coherently merging two optical pulses from different spectral regions. The shape of the optical waveform is controlled via two synthesis parameters, the relative-phase $\varphi$ (RP) among two portions of the synthesized optical spectrum (i.e., the two channels of our parametric waveform synthesizer) and the overall carrier-envelope phase $\psi$ (CEP). By scanning the two phases, we can produce in a reproducible way a large set of waveforms and record the HHG spectra corresponding to each ($\varphi_{i}$,$\psi_{i}$) pair. In addition, attosecond streaking measurements were conducted for selected waveforms to reveal the temporal structure of the underlying isolated attosecond pulses. Simulations of the HHG process qualitatively support our key experimental findings and allow us to better understand the impact of the optical waveform on the IAP-characteristics.\\

\section*{Tailored driver fields via parametric waveform synthesis}
\begin{figure}
\includegraphics[width=\textwidth]{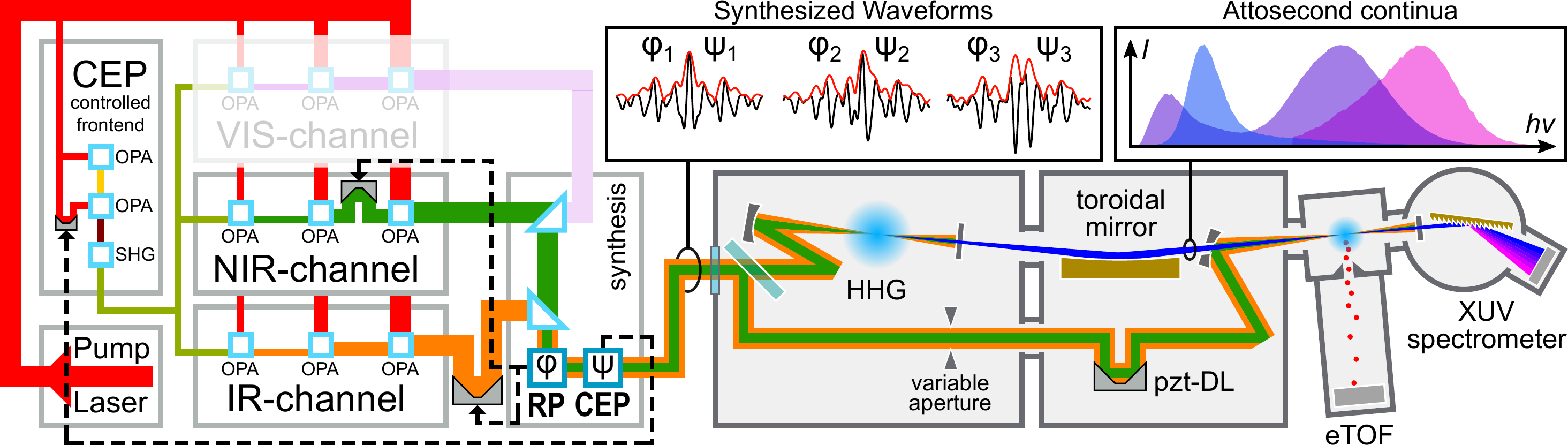}
%\internallinenumbers
\caption{\small{\textbf{Experimental setup.} (on the left) Parametric waveform synthesizer, consisting of a cryogenic Ti:Sapphire pump laser, a CEP-controlled seeding frontend, two spectral channels (NIR and IR) and a multi-phase meter that allows to measure the RP ($\varphi$) among the spectral channels and the CEP ($\psi$) of the synthesized output. A feedback system acting on multiple delay-lines allows to stabilize and control ($\varphi$, $\psi$), and therefore the output waveforms. The waveforms entering the attosecond beamline (on the right) are focused into an HHG gas target. The generated HHG radiation is separated from the driving optical field via a metallic filter and refocused by an Au-coated toroidal mirror into the electron collecting volume of a Kaesdorf ETF11 electron time-of-flight spectrometer, used for attosecond streaking measurements. The spectra of the HHG emission are measured by a McPherson 251MX grating spectrometer, equipped with an Andor Newton 940 CCD camera.}}
\label{Fig:Setup}
\end{figure}

The optical waveforms utilized in our experiments are generated by our recently demonstrated parametric waveform synthesizer (PWS)\cite{rossi2020sub}. The PWS is based on optical parametric amplifiers (OPAs, see Fig.~\ref{Fig:Setup}), pumped by a cryogenic Ti:Sapphire laser amplifier, where different spectral regions of a multi-octave spanning CEP-controlled seed pulse are individually generated, amplified in three consecutive OPA-stages and coherently recombined after compression. The two spectral channels employed in this experiment, namely a near-infrared (NIR) and infrared channel (IR), span the spectral regions of 650-1000 nm and 1200-2200 nm, and support pulses with full-width half-maximum (FWHM) durations of $\sim$6\,fs and $\sim$8\,fs and with pulse energy up to 0.15\,mJ and 0.6\,mJ respectively. The 1.7\,octave-spanning waveforms can be shaped by acting on the RP $\varphi$ (or relative delay) among the NIR and IR pulses and the CEP $\psi$ of the combined pulse. The relative phase $\varphi$ can be locked and controlled over a range of $\sim$200\,rad ($\sim$100\,fs) with residual noise of $<$0.1\,rad rms, where the CEP $\psi$ can be varied over a $\sim$10\,rad range with residual noise of $<$0.25\,rad rms. For $\varphi=0$ , where the temporal envelopes of NIR and IR pulses overlap, the FWHM duration of the synthesized waveform reaches below 3\,fs, that corresponds to less than one optical cycle (sub-cycle duration).
The control system that stabilizes the RP and CEP jointly with multi-beam pointing stabilization ensure stable and reproducible synthesis over many hours\cite{silva2020waveform}.\\
Inside the attosecond beamline (Fig.~\ref{Fig:Setup}), the synthesized pulses are split into two arms, the more intense replica drives the HHG, the other one is used as streaking-field. The pulse energy of each channel at the HHG interaction point measures 70\,$\mu$J (NIR) and 170\,$\mu$J (IR), resulting in a synthesized waveform whose most intense cycle has a half-period (distance between neighboring electric field zero crossings) ranging from 1.3\,fs to 3.2\,fs depending on the RP and CEP setting. Such waveforms, when focused by a spherical mirror ($f$ = 375\,mm) into a $\sim$ 2\,mm long gas cell, reach a maximum peak intensity (for $\psi$ = 0, $\varphi$ = 0) of 2-3$\times$10$^{15}$\,W/cm$^{2}$ in the finite HHG gas cell. These conditions apply to all experimental data presented here.\\
\newpage
\section*{Waveform-dependent attosecond continua}
\begin{figure}
\includegraphics[width=\textwidth]{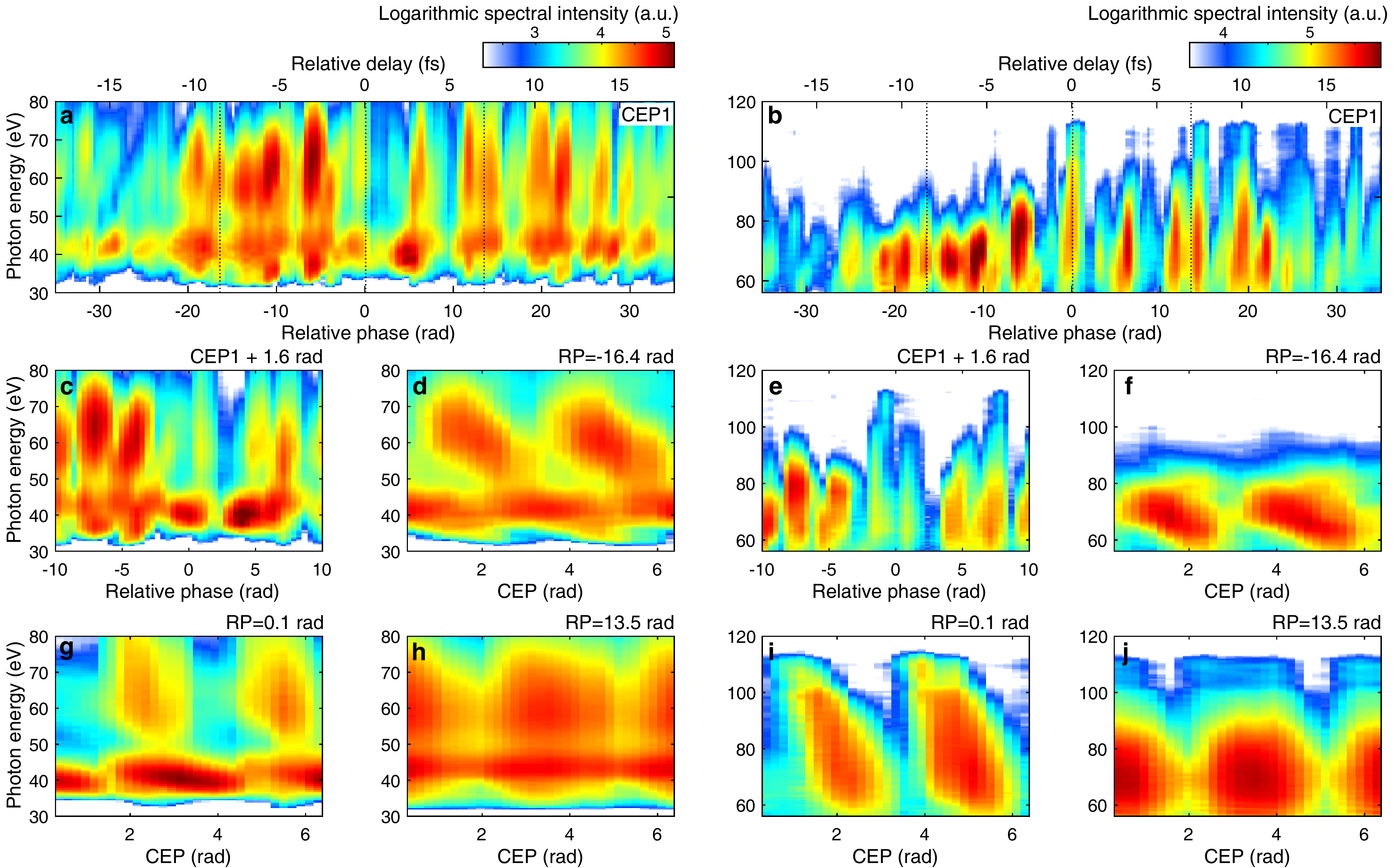}
%\internallinenumbers
\caption{\small{\textbf{Dependency of HHG spectra on RP and CEP in argon.} The HHG process is driven in argon by the synthesized pulses and the XUV spectra are recorded with a grating XUV spectrometer. Depending on the spectral range, the spectra are dispersed by either a grating with 300\,groove/mm (low photon energy) (\textbf{a, c, d, g, h}) or with 1200\,groove/mm (high photon energy) (\textbf{b, e, f, i, j}). With the CEP fixed, XUV-spectra vary significantly with changing RP (\textbf{a, b}), where negative relative phase means the IR pulse arrives first. A CEP shift by $\pi$/2 completely changes the XUV spectrum (\textbf{c, e}). On the other hand, XUV-spectra vary with CEP and the landscape of the CEP-dependent variation appears different when the RP is fixed at another value. (\textbf{d, f}), (\textbf{g, i}) and (\textbf{h, j}) correspond to the XUV spectrum CEP-dependent variation with the RP fixed at 13.5\,rad, 0.1\,rad and -16.4\,rad, marked with dashed line in (\textbf{a, b}), respectively. The integration time for each spectrum is 200\,ms and thin metallic filters are applied to block the driving laser beam (500\,nm thick beryllium).}}
\label{Fig:HHGAr}
\end{figure}

\begin{figure}
\includegraphics[width=\textwidth]{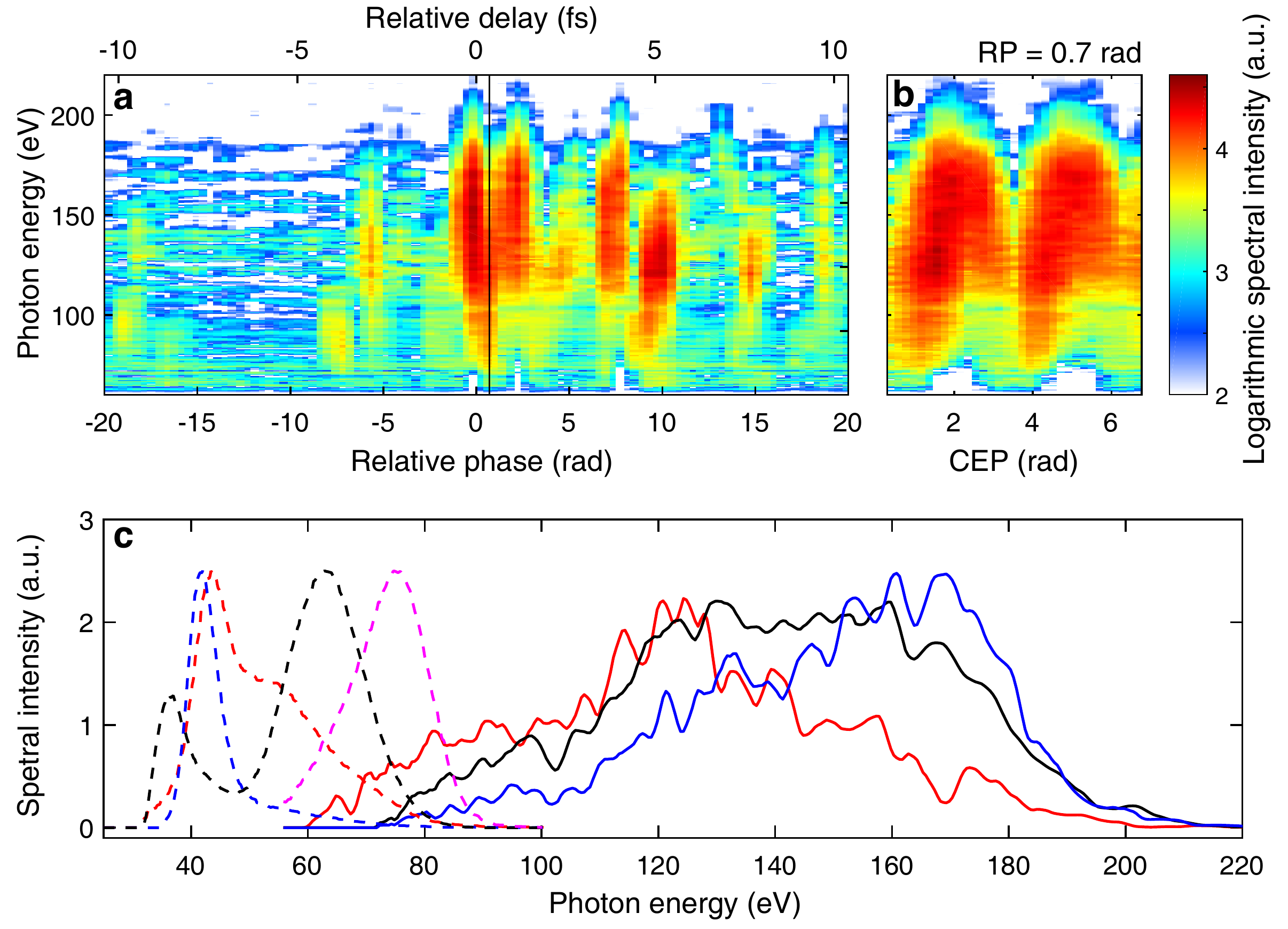}
%\internallinenumbers
\caption{\small{\textbf{Dependency of HHG spectra on RP and CEP in neon.} The XUV spectrum is dispersed with a 1200\,groove/mm grating. With the CEP fixed, XUV spectra vary by changing the RP \textbf{a.} CEP-dependent XUV spectra at fixed RP = 0.7\,rad, marked with a black line in \textbf{a}, is shown in \textbf{b}. Four spectra of HHG in argon (dashed, from Fig. \ref{Fig:HHGAr}) and three spectra of HHG in neon (solid) are presented to showcase the spectral tuning range. \textbf{c}. The integration time for each neon HHG spectrum is 200\,ms and a Zr filter (300\,nm) is used to block the driving laser beam.}}
\label{Fig:HHGNe}
\end{figure}

To fully showcase the variety of HHG emissions that results from controlling the driving waveform, HHG spectra were recorded for a large set of waveforms, which were obtained by systematically scanning through both RP and CEP. The following measuring procedure was used: First, we concentrated on the generation of harmonics in argon and their detection in the range 30-70\,eV (Fig.~\ref{Fig:HHGAr}). After determining the conditions favourable for harmonic generation ($\sim$300 mbar of pressure, gas-cell positioned a few mm after the focus), we started the waveform scanning procedure. Initially the control system moved the RP towards a given negative value (IR pulse ahead of NIR pulse). Then the CEP was scanned periodically ($T$ = 4\,s) with a linear ramp function over a range of $\sim$10\,rad. After two complete CEP scans, the RP was incremented step-wise (with 0.8\,rad steps). The procedure was repeated until a certain positive value of RP (IR pulse behind NIR pulse) was reached. Once the measurement was completed, a resorting algorithm rearranged all HHG-spectra into an N$\times$M matrix, according to the corresponding ($\varphi_i,\psi_i$) pair. The RP bin-size is choosen to be equal to its step-size during the scan. The CEP instead is a linear function and the CEP-bin size was chosen as a compromise between CEP-resolution and redundancy/complteness of the data set. This approach allows us to obtain a complete data set of 100$\times$16 HHG-responses (RP/CEP steps) with 3-10 HHG spectra per matrix element in less than an hour (see Supplementary Information).\\
Since the resulting four-dimensional matrix (HHG photon-energy, spectral intensity, RP and CEP) cannot be visualized directly, we present, in Fig.~\ref{Fig:HHGAr}, cuts through the data-set for either a fixed RP or fixed CEP value. In Fig.~\ref{Fig:HHGAr}(a) and (b) it is possible to observe the rich tuning potential of the HHG spectra that can be generated by changing the RP while keeping the CEP constant. One can observe that almost all spectra are free of spectral-fringes and therefore likely correspond to isolated attosecond pulses (as we will verify in the next section). This result, which may seem surprising at first glance, is explained by the asymmetric and non-sinusoidal nature of the synthesized waveforms. In fact, although the waveforms can reach a sub-cycle FWHM duration only for RP-values close to 0, the superposition between the two channels leads to a single field oscillation being significantly higher than the others for a much wider range of RP values. In addition to this, there is a unique property of these extremely broadband waveforms, where each optical half-cycle exhibits a unique, non-sinusoidal field shape that is different from the preceding and subsequent ones. For example, the duration of such half-cycle can vary between 1.3\,fs and 3.2\,fs, greatly affecting the electron excursion time and, therefore, the probability of HHG emission. Furthermore, the exact shape of the electric field contained in a certain portion of the cycle will favour specific electron trajectories, resulting in a wide range of observed HHG spectra. The attosecond continua generated in argon span the 35-110\,eV range. Narrowband continua with different photon-energies can be found, for instance for $\varphi$ = -28\,rad ($\sim$5\,eV bandwidth peaked at 42\,eV) or $\varphi$ = -16\,rad ($\sim$15\,eV bandwidth peaked at 60\,eV), as well as broadband continua such as the one around $\varphi$ = 0\,rad (spanning 1.5 octaves, from $\sim$36\,eV to $\sim$101\,eV, peaked at 75\,eV). Interestingly the highest cut-off energy, occurring around $\varphi$ = 0\,rad, does not coincides with the highest yield region, that instead appears around $\varphi$ = -5\,rad, suggesting that the additional control obtained by shaping the waveform may also help improving HHG conversion efficiency beyond what is possible by pure pulse compression.\\
When changing the CEP by $\pi$/2, the landscape of the RP-dependent XUV-continua changes completely as shown in Fig.~\ref{Fig:HHGAr}(c) and (e), where for instance the HHG yield greatly reduces around $\varphi$ = 0 when compared to Fig.~\ref{Fig:HHGAr}(a, b). In Fig.~\ref{Fig:HHGAr}(d, e), (g, i) and (h, j) instead we plot the CEP dependency for three different RP values. For all three RP values we observe the typical CEP dependency of HHG continua in the cut-off region, but here extended to the entire observed spectral range. This verifies that IAPs can be generated over broad spectral ranges without the need for additional gating techniques, such as double optical gating\cite{chang2007controlling} or by filtering the cut-off emission\cite{goulielmakis2008single} (amplitude gating), but shaped solely by choosing an appropriate optical waveform. Moreover, it is interesting to notice how the three continua evolve differently when changing the CEP: in (d,f) the overall intensity drops without a significant change in spectral shape; in (g,i) instead the change in CEP is accompanied by a central photon-energy shift; finally in (h,j) the central energy does not vary considerably but the spectrum broadens and forms a double-hump structure. These observations suggest the vastness of the possibilities introduced by electron trajectory coherent control via waveform synthesis.\\
We repeated a similar RP and CEP scan in neon to reach higher photon energies. All the macroscopic parameteres stayed the same as in the previous case, including the gas pressure ($\sim$300\,mbar), only the gas target position was optimized (moved closer to the laser focus). Remarkably, the neon HHG was sufficiently intense to be properly detected without increasing the integration time of the spectrometer when compared to the argon case (200\,ms).\\
Fig.~\ref{Fig:HHGNe} shows that the tunability of attosecond continua can be extended into the soft-X ray region up to $\sim$200\,eV in neon. Controlling the RP (see Fig.~\ref{Fig:HHGNe}(a)) allows once more to tune the central photon-energy and the bandwidth, as highlighted in Fig.~\ref{Fig:HHGNe}(c) where three exemplary spectra are extracted from Fig.~\ref{Fig:HHGNe}(a) and are centred at 120\,eV, 145\,eV and 170\,eV respectively. In Fig.~\ref{Fig:HHGNe}(c), several exemplary spectra of HHG in argon (dashed line) are also shown, joining the neon HHG spectra, to showcase the wide tuning range of the generated HHG continua. The CEP dependency shown in Fig.~\ref{Fig:HHGNe}(b) supports that the continua observed here corresponds to IAPs as well.\\
The waveform control demonstrated here allows generating attosecond continua with tunable central energy and bandwidth over an impressive 30-200 eV range simply replacing the gas type and optimizing the gas target position without modifying any other macroscopic HHG parameters.
\\
\newpage
\newpage
\section*{Tunable isolated attosecond pulse characterization}
\begin{figure}
\centering
\includegraphics[width=14cm]{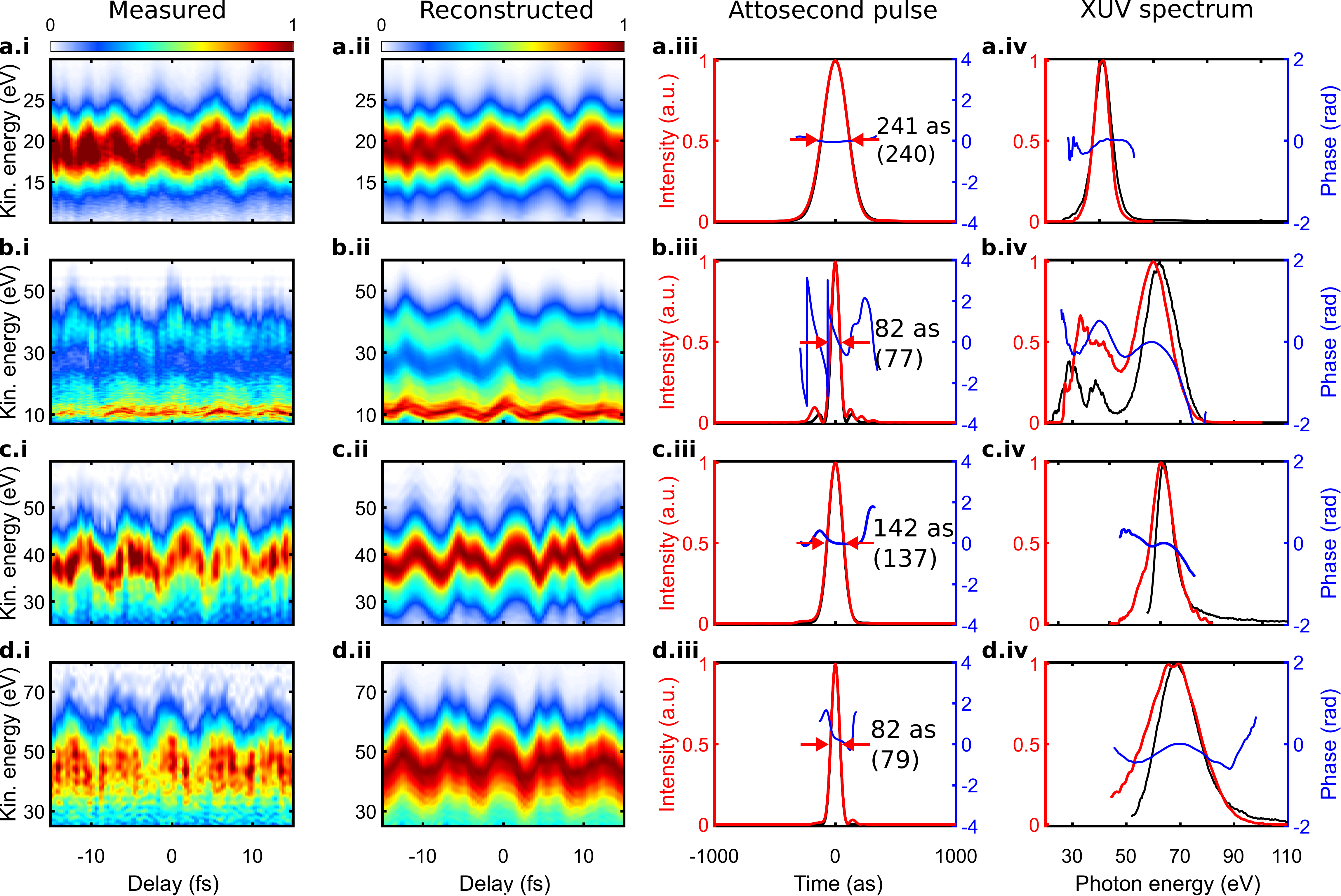}
%\internallinenumbers
\caption{\small{\textbf{Isolated attosecond pulses characterized via attosecond streaking.} The photoelectron spectrograms \textbf{a-d}(i) are obtained with HHG from argon (\textbf{a, b}) and neon (\textbf{c, d}). The VTGPA reconstructed traces \textbf{a-d}(ii) are shown along with to the experimental traces \textbf{a-d} (i). The retrieved temporal profiles (red) and temporal phase (blue) of the IAPs are shown in \textbf{a-d}(iii). The retrieved and the Fourier transform limited (FTL, bracket) pulse duration are both stated (iii). The retrieved XUV spectrum (red) and spectral phase (blue) are shown in \textbf{a-d}(iv). The independently measured XUV spectra (black) (with grating XUV spectrometer) are also shown for each case (iv) to compare with the retrieved ones.}}
\label{Fig:IAP}
\end{figure}

So far, we have characterized the spectral magnitude of the attosecond continua, but to gain insights on the temporal profile of the IAPs it is necessary to perform attosecond streaking measurements. Here, the HHG beam overlaps with the optical laser pulse at a gas target and the momenta of the HHG-ionized photoelectrons are modulated by the optical streaking field. The photoelectron spectrogram is acquired by varying the delay between the optical streaking field and the HHG pulses. By feeding the measured spectrogram to an iterative phase-retrieving algorithm, the spectrotemporal information of HHG radiation and of the streaking field, can be obtained.\\
Given the complexity of our streaking waveforms and the broad bandwidth of the generated attosecond continua, we decided to use the Volkov Transform Generalized Projection Algorithm (VTGPA)\cite{keathley2016volkov} for the pulse retrieval. VTGPA has advantages over conventional attosecond pulse reconstruction methods\cite{mairesse2005frequency, chini2010characterizing,laurent2013attosecond} as it does not rely on the central-momentum approximation, therefore allowing us to precisely retrieve broadband XUV spectra, and it has better immunity to detection noise. Moreover, the VTGPA code is ideal for achieving a correct retrieval in the presence of complex streaking waveforms since it is less restrictive on the intensity and bandwidth of the streaking field. As the data-acquisition of a single streaking trace can take up to a couple of hours, it is only feasible to conduct such extensive characterization for a few selected cases, as shown in Fig.~\ref{Fig:IAP}.\\
All attosecond streaking retrievals confirm that the observed HHG continua indeed correspond to isolated attosecond pulses. Moreover, these measurements show that the demonstrated tunability extends also to the attosecond pulse duration. In the argon case, the IAP in Fig.~\ref{Fig:IAP}a(iii) shows a duration of 240\,as, while the IAP in Fig.~\ref{Fig:IAP}b(iii) only 80\,as. This control over the IAP duration also applies to the case of neon, where different ($\varphi_{i}$,$\psi_{i}$) pairs lead to different continua bandwidths, and the attosecond pulse duration decreases from 140\,as to 80\,as. Here it is worth noticing that the measured photoelectron spectrograms a-d(i) and the reconstructed ones a-d(ii) not only match each other well, but are also in substantial agreement with the independently measured XUV spectra, demonstrating the fidelity of our measurements. These measurements showcase that, by shaping the HHG-driving waveform, we can not only control the spectral properties of the IAPs, but also gain control on their pulse durations. This represents a significant advance in attosecond source technology, since different gating techniques can either deliver narrowband or broadband IAPs, but do not allow to easily switch between them\cite{chini2014generation}.\\
\newpage
\section*{HHG simulations with tailored waveforms}
\begin{figure}
\centering
\includegraphics[width=0.9\textwidth]{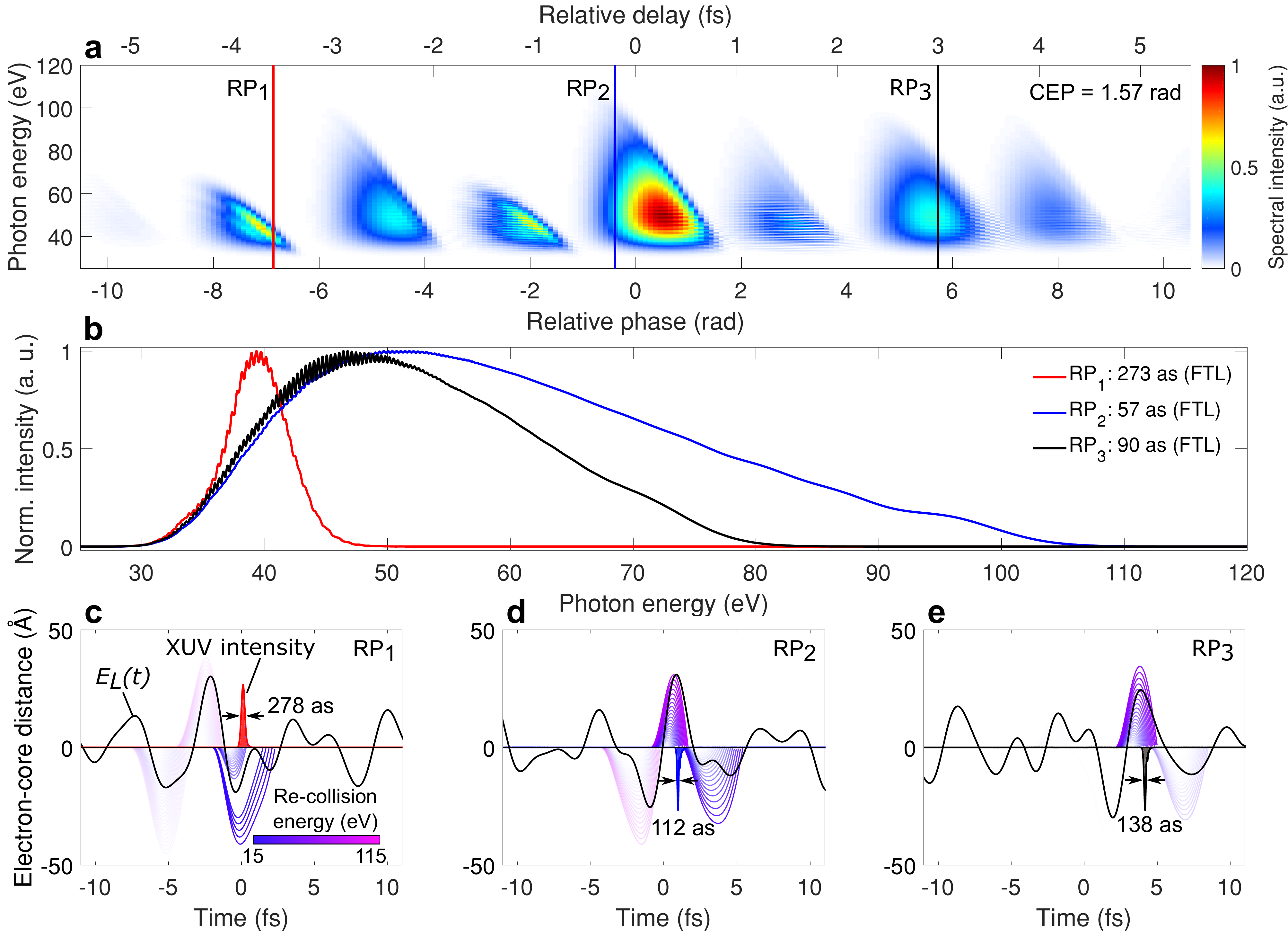}
%\internallinenumbers
\caption{\small{\textbf{HHG single atom response simulations for argon.} \textbf{a}, Dependence of XUV spectrum on RP between the NIR and IR channels, using a common CEP fixed at $\pi$/2. The Lewenstein integral\cite{lewenstein1994theory} considers short trajectories only and is solved using synthesized driving fields $E_{L}(t)$ built upon the combination of the actual NIR and IR channel temporal profiles of the PWS. The synthesized peak intensity was fixed to 2.5$\times$10$^{14}$W/cm$^{2}$ for RP = 0 and CEP = 0. Linear 1D propagation of the resulting single-atom XUV pulse through the gas target (2\,mm, 300\,mbar) is implemented. \textbf{b}, Lineouts of the XUV spectra at different RP values and their corresponding FTL pulse durations. \textbf{c-e}, classical electron trajectories (purple) and emitted XUV attosecond bursts (shaded areas) are shown for different optical waveforms (black solid lines) resulting from the RP values indicated in \textbf{a}. The trajectory color-coding is based on the re-collision energy and the transparency level is based on the ionization rate at the trajectory birth time (see Supplementary Information).}}
\label{Fig:SimHH}
\end{figure}

Attosecond pulse generation driven by complex waveforms was also investigated theoretically by means of classical trajectory analysis and quantum single-atom response simulations. We started by calculating the HHG obtained during a RP-scan (with fixed CEP) similar to the experimental RP-scan shown in Fig.~\ref{Fig:HHGAr}. In the calculations we started with the individual NIR and IR pulses of our PWS characterized via two-dimensional shearing spectral interferometry (2DSI\cite{birge2006two}, see Supplementary Information). Then we calculated the overall synthesized field for RP values between -11\,rad to 11\,rad. Finally, we use these waveforms as HHG driving field in a single-atom response simulation (based on HHGMax\cite{hhgmax}, see Fig.~\ref{Fig:SimHH}a). Despite the quantitative differences observed between experiment and simulation results, the main features observed experimentally are reproduced by the simulation. In particular, we observe the overall asymmetry of the RP-scan (RP$>$0 vs.$<$0) as well as the periodic appearance of bright HHG-continua alternating with low-yield regions. For certain RP values the continua appear to be broadband (close to $\varphi$ = 0, where the envelope peak of the NIR and IR pulses overlap) while for other values narrowband ($\varphi$ = -6.8 rad) or with intermediate bandwidth ($\varphi$ = 5.7 rad). Moreover, the HHG spectra exhibit spectral-fringes only for few RP values and the fringes are located in the low photon-energy region, similar to the experimentally observed ones.\\
We then analyse the three particular waveforms corresponding to these continua, one leading to a narrowband continuum (Fig.~\ref{Fig:SimHH}c), one to a broadband one (Fig.~\ref{Fig:SimHH}d) and the third to an intermediate bandwidth (Fig.~\ref{Fig:SimHH}e). In the first case the half-cycle peaked at $t$ = -2\,fs is by far the most intense one and therefore the only significant contributor to strong-field ionization. The following half-cycle is weaker and the accelerated electrons only gain a moderate amount of energy, leading to narrowband IAPs with low photon-energy and a duration of 278\,as, when accounting for the dispersion due to the 2\,mm thick, 300\,mbar, neutral gas column. The IAP spectrum and duration are in relatively good agreement with the narrowband IAP presented in Fig.~\ref{Fig:IAP}a(iii). The second waveform instead leads to a broadband emission. Here an intense half-cycle peaked at $t$ = -1\,fs ionizes a certain fraction of the atoms. The electrons are then accelerated by the following half-cycle, in this case more intense than the preceding one. As underlined by the classical trajectory analysis, these electrons, depending on their precise ionization time, gain significant momentum leading to a broadband emission that extends to higher photon-energies than in the previous case. Given that the accelerating half-cycle also leads to a significant strong-field ionization, one would expect the appearance of a second attosecond pulse. Interestingly however, the following half-cycle (peaked at $t$ = 4.7\,fs) has a significantly longer period and lower field amplitude, which may dramatically reduce the emission probability. This reduction in emission probability leads to the suppression of the second attosecond pulse and, therefore, to the appearance of a well-isolated broadband attosecond pulse with a duration of 112\,as. For the third waveform, which falls between the two previous cases, the cycle that provides the dominant contribution to the HHG arrives after $t$ = 0. Electrons contributing to HHG originate mainly from the half-cycle peaked at $t$ = 2\,fs, since the half-cycle just before $t$ = 0 is too weak to significantly ionize the atoms. The field strength of the accelerating sub-cycle is intermediate, which leads to a less extended cut-off energy.\\
These simulations, though far from being exhaustive, allow us to shed light on some of the fundamental mechanisms, which allow us to generate tunable IAPs by controlling the optical waveform via the pulse synthesis parameters.\\
\newpage
\section*{Conclusion}
We have experimentally demonstrated that the spectrotemporal characteristics of isolated attosecond pulses can be controlled in manifold ways by shaping the sub-cycle features of the broadband HHG driving pulses. The capability to shape the waveform was achieved by adjusting both the RP among two halves of the optical spectrum and the total CEP of the synthesized field. For the similar macroscopic HHG-conditions, we generate continua across the 30-110 eV range in argon and extend it up to 200\,eV in neon. The bandwidth and duration of the generated attosecond pulses are substantially varied solely by changing the RP and CEP settings. This unprecedented spectral tunability for an attosecond source is achieved without the use of any additional gating or spectral filtering technique. HHG simulations help elucidating some of the underling physical mechanisms, unambiguously linking the IAP tunability with the shape of the HHG-driving waveform. We believe that parametric waveform control will allow attosecond light sources to leap into a new era of development, with enormous benefits for attosecond science experiments.\\
The strong-field coherent control experiments presented here offer a glimpse of upcoming technologies enabled by waveform-controlled light-matter interactions.\\
\newpage
\bibliographystyle{naturemag}
\bibliography{nat.phot.iaps}
\newpage
\begin{addendum}
\item[Acknowledgement] 
We gratefully acknowledge support from Deutsches Elektronen-Synchrotron (DESY), a Center of the Helmholtz Association, the Cluster of Excellence ’CUI: Advanced Imaging of Matter’ of the Deutsche Forschungsgemeinschaft (DFG) - EXC 2056 - project ID 390715994, the priority programme ’Quantum Dynamics in Tailored Intense Fields’ (QUTIF) of the DFG (SPP1840 SOLSTICE) and the PIER Hamburg-MIT Program. We thank A. Trabattoni for many valuable discussions and O.D. Mücke for initial work on the PWS. P. Keathley acknowledges support by the Air Force Office of Scientific Research (AFOSR) Grant under Contract No. FA9550-19-1-0065.
\item[Author contributions] Y.Y. designed and implemented the attosecond beamline. G.M.R. and R.E.M. designed and built the opto-mechanical set-up and its waveform stabilization/control infrastructure. Y.Y., R.E.M., G.M.R., F.S. and M.A.S.-T conducted the experiments. F.S. and M.A.S.-T carried out the data processing and its analysis jointly with R.E.M. and P.D.K. The numerical simulations were performed by M.A.S.-T. The original idea of the experiment was conceived by Y.Y., R.E.M., G.M.R and F.X.K.. Y.Y., R.E.M. and G.M.R. co-wrote the paper with contributions from all authors. The project was supervised by G.C. and F.X.K.
\item[Data and code availability]
The experimental data and computer codes used in this paper are available from the corresponding author upon reasonable request.
\end{addendum}

\end{document}